\documentclass[twocolumn]{aastex62}
\usepackage{xcolor}
\usepackage{pifont}
\newcommand{\cmark}{\ding{51}}%
\newcommand{\xmark}{\ding{55}}%

\definecolor{darkbrown}{HTML}{8c4600}
\definecolor{darkerbrown}{HTML}{501010}
\definecolor{darkblue}{HTML}{1833a1}
\newcommand{\treinf}{t_{\rm reinf}}

\newcommand{\vk}{v_{\rm k}}

\newcommand{\Omegainv}{\Omega^{-1}}
\newcommand{\Sigmag}{\Sigma_{\rm g}}
\newcommand{\Sigmap}{\Sigma_{\rm p}}

\newcommand{\cs}{c_{\rm s}}
\newcommand{\rhos}{\rho_{\rm s}}
\newcommand{\rhop}{\rho_{\rm p}}
\newcommand{\rhog}{\rho_{\rm g}}

\graphicspath{{./}{plots/}}

\submitjournal{ApJ}

\shorttitle{Resilience of Planetesimal Formation in Weakly-Reinforced Pressure Bumps}
\shortauthors{Carrera et al.}

\begin{document}

\title{Resilience of Planetesimal Formation in Weakly-Reinforced Pressure Bumps}


\correspondingauthor{Daniel Carrera}
\email{dcarrera@gmail.com}

\author[0000-0001-6259-3575]{Daniel Carrera}
\affiliation{Department of Physics and Astronomy, Iowa State University, Ames, IA, 50010, USA}

\author{Andrew J. Thomas}
\affiliation{Department of Physics and Astronomy, Iowa State University, Ames, IA, 50010, USA}

\author[0000-0002-3771-8054]{Jacob B. Simon}
\affiliation{Department of Physics and Astronomy, Iowa State University, Ames, IA, 50010, USA}

\author{Matthew A. Small}
\affiliation{Department of Physics and Astronomy, Iowa State University, Ames, IA, 50010, USA}

\author{Katherine A. Kretke}
\affiliation{Department of Space Studies, Southwest Research Institute, Boulder, CO 80302, USA}

\author[0000-0002-8227-5467]{Hubert Klahr}
\affiliation{Max-Planck-Institut f{\"u}r Astronomie: Heidelberg, Baden-W{\"u}rttemberg}

\begin{abstract}
The discovery that axisymmetric dust rings are ubiquitous in protoplanetary disks has provoked a flurry of research on the role of pressure bumps in planet formation. High-resolution simulations by our group have shown that even a modest bump can collect enough dust to trigger planetesimal formation by the streaming instability.
In this work, we probe the limits of planetesimal formation when the external source of pressure bump reinforcement is extremely weak.
We conduct simulations of radially elongated shearing boxes to capture the entire bump, which is generated and maintained over some timescale $\treinf$ by a Newtonian relaxation scheme.
We find that planetesimal formation is extremely resilient for cm-sized grains. We reduced the strength of reinforcement by up to a factor of 100 and the location and initial masses of planetesimals were essentially unaffected.
However, we do find that strong reinforcement causes much faster pebble drift compared to the the standard pebble drift rates. The resulting larger pebble flux enhances the planetesimal growth rate by pebble accretion. We hypothesize that to sustain the bump, our code has to extract angular momentum (the strength of this negative torque depends on $\treinf$), and some of this torque is transferred to the particles, causing them to drift faster for a stronger torque (i.e., smaller $\treinf$). Since any physical process that sustains a pressure bump must do so by torquing the gas, we conjecture that the effect on pebble drift is a real phenomenon, motivating further work with physically realistic sources to generate the bump.
\end{abstract}


%
%
\section{Introduction}
\label{sec:intro}

The now iconic image of the dust continuum emission around HL Tau \citep{Alma_2015} and subsequent ALMA observations \citep[e.g.][]{Huang_2018} have shown that axisymmetric dust rings are a very common feature of young planet-forming disks. The most popular explanation is that these rings are produced by pressure bumps \citep[e.g.,][]{Dullemond_2018a}, possibly caused by massive planets \citep[e.g.,][]{Picogna_2015,Fedele_2017,Hendler_2018,Baruteau_2019,Perez_2019,Toci_2020,Veronesi_2020}. However, the frequency of these structures seems to greatly exceed the known frequency of giant planets \citep{Ndugu_2019}. Pressure bumps can be produced by other mechanisms such as MHD zonal flows \citep[e.g.][]{Kunz_2013,Flock_2015,Bethune_2016}, evaporation fronts \citep[e.g.][]{Brauer_2008}, disk winds \citep[][]{Suriano_2018,Riols_2019,Riols_2020}, thermal waves induced by stellar irradiation \citep{Dullemond_2000,Watanabe_2008,Ueda_2021,Wu_2021}, or a dust-driven inhibition of MRI turbulence \citep[][]{Dullemond_2018b}. Aside from pressure bumps, dust rings may also form by sintering of icy aggregates \citep{Okuzumi_2016} or through various gravitational instabilities \citep[][]{Takahashi_2016,Tominaga_2019}. In addition, \citet{Ohashi_2021} showed that very young Class 0/I disks ($t_{\rm disk} \sim 10^4$~yr) may also develop a dust ring as a by-product of the early phases of dust evolution, as the dust coagulation front sweeps outward through the disk.

In this work we investigate the potential role of pressure bumps in planet formation. Since the mechanisms for creating pressure bumps are still not well understood (and in fact, different bumps could be produced by different mechanisms), one of the goals of our investigation is to better understand how the source of the bump affects planet formation in a model-agnostic manner.

However they form, pressure bumps may be natural sites for the formation of planetesimals. Planetesimals are 1--100 km bodies that are the building blocks of rocky planets and giant planet cores. The most compelling model is that they form when an local enhancement of the dust-to-gas ratio of around $Z \sim 0.02 - 0.04$ \citep{Carrera_2015,Yang_2017,Li_2021} triggers a runaway concentration of dust known as the streaming instability (SI) \citep{Youdin_2005,Youdin_2007,Squire_2020}. The SI collects solid particles into azimuthal filaments \cite[e.g.,][]{Johansen_2007a,Johansen_2007b,Bai_2010a,Carrera_2015,Yang_2017} that can reach very high densities. If the particle density reaches the Roche density,

\begin{equation}
    \rho_{\rm roche} = \frac{9\Omega^2}{4\pi G},
\end{equation}
where $G$ is the gravitational constant and $\Omega$ is the orbital frequency, then particle self-gravity overpowers tidal forces (Keplerian shear) and the particles form self-gravitating particle clumps \citep{Simon_2016,Simon_2017,Abod_2019,Li_2019,Gerbig_2020,Carrera_2021}. From that point, the gravitational instability (GI) leads to rapid collapse into solid planetesimals \citep{Goldreich_1973,Nesvorny_2010,Jansson_2014}. Perhaps the most important feature of the GI is that it bypasses the growth barriers that seem to impede a gradual collisional growth scenario, such as the bouncing and fragmentation barriers \citep[e.g.,][]{Guttler_2010,Zsom_2010}, the radial drift barrier \citep[e.g.,][]{Weidenschilling_1977,Birnstiel_2012}, and the recently proposed ablation barrier \citep{Rozner_2020}. In terms of observations, a kink in the size distribution of asteroids at around $\sim 100$km is evidence that asteroids formed from the GI collapse of a pebble cloud \citep{Morbidelli_2009}. In addition, the angular momentum distribution of trans-Neptunian binaries is consistent with that of pebble clouds produced by the SI \citep{Nesvorny_2019}.

The critical link between planetesimal formation and ALMA dust rings is that the runaway concentration of solids caused by the SI only takes place if the dust-to-gas ratio $Z$ is already enhanced relative to the $Z \sim 0.01$ value of the interstellar medium and the solar nebula, at least in the small particle regime \citep{Carrera_2015,Yang_2017,Li_2021}. The exact particle size where the SI requires $Z > 0.01$ is uncertain because the SI is also less efficient when the radial pressure gradient is steeper \citep{Bai_2010b}, as is the case in our investigation. In any case, ALMA dust rings are local enhancements of $Z$ caused by the aerodynamic concentration of solids induced by axisymmetric pressure bumps. The notion that pressure bumps may act as SI-driven planetesimal factories has sparked multiple investigations \citep[e.g.][]{Taki_2016,Onishi_2017,Lenz_2019,Carrera_2021}. For example, \citet{Carrera_2021} ran the first fully 3D simulations of the SI in a domain large enough to contain a pressure bump with scales comparable to observed disk rings. They found that even a small bump can trigger the SI and a particle trap (i.e., halting inward particle drift) is not needed to efficiently convert dust into planetesimals.

With these results in hand, we are now in a position to explore different kinds of pressure bumps (at least within the constraints of our numerical setup). The primary question we aim to address is whether or not planetesimal formation in pressure bumps is robust to different reinforcement strengths.  As alluded to above, planets will reinforce bumps on a short (i.e., dynamical) timescale.  However, unless the very first planetesimals formed in a completely different manner that did not require pressure bumps, the very first bumps must have been produced by something other than planets. As such, these bumps would not necessarily be reinforced at the same strength as those induced by planets.   Thus, in this paper we present the fully 3D simulations of planetesimal formation by the SI in pressure bumps with weak reinforcement. Our results will test the general viability of planetesimal formation in bumps produced by something other than planets.

This paper is organized as follows. In section \S \ref{sec:methods} we describe our numerical methods and initial conditions. In section \S \ref{sec:results} we present our results,. Finally, we summarize and draw conclusions in section \S \ref{sec:conclusions}.

%
%
\section{Methods} 
\label{sec:methods}

We employ identical methods, parameters, and initial conditions to \cite{Carrera_2021}, except that our runs trade lower resolution for longer simulation times, and we vary the values of the pressure bump amplitude, reinforcement time $\treinf$, and the vertical extent of the simulation $L_z$. For this reason, we only give a brief description of our numerical methods and refer the interested reader to \cite{Carrera_2021} for more details.

\subsection{Numerical Algorithm}
\label{sec:methods:numerical}

We carry out a series of local, shearing box (i.e., co-rotating disk patch) simulations with the {\sc Athena} code \citep{Stone_2008}, a second-order accurate, flux-conservative Godunov code for solving the gas dynamic equations in astrophysical environments. We employ the code's hydrodynamic mode (i.e., no magnetic fields) and include particles, accounting for gas drag on particles, particle feedback on the gas, and particle self-gravity.  We assume an isothermal gas and no externally imposed turbulence. As in previous works, we employ an inward force acting on the particles to account for radial drift (see \citealt{Bai_2010a}) and a particle-mesh Fast Fourier Transform for including particle self-gravity. For more code details and tests, both in terms of general applications and specifics to our setup, see \cite{Stone_2008,Stone_2010,Bai_2010a,Simon_2016,Li_2018}. 

Our simulation domain consists of a long and narrow simulation domain with $L_x \times L_y = 9H\times 0.2H$, where $H$ is the gas vertical scale height (the box height $L_z$ varies between runs). All simulations have a resolution of $320/H$, and the total number of particles is equal to the total number of grid cells. As in \cite{Carrera_2021}, our setup is based on a disk model as follows. The disk has mass $M_{\rm disk} = 0.09 M_\odot$ around a solar-mass star with surface density

\begin{equation}
    \Sigma(r) = \frac{M_{\rm disk}}{2\pi r_c} r^{-1}
\end{equation}
where $r_c = 200$AU. We also assume an optically thin disk with temperature profile

\begin{equation}
    T(r) = 280{\rm K} \left( \frac{r}{{\rm AU}} \right)^{-1/2}
\end{equation}
With this temperature profile, we get a slightly flared disk with aspect ratio

\begin{equation}
    \frac{H}{r} = \frac{\cs}{\vk} = 0.033 \left( \frac{r}{{\rm AU}} \right)^{1/4}
\end{equation}
where $H$ is the disk scale height, $\cs$ is the sound speed, and $\vk$ is the Keplerian orbital speed. All simulations begin with a vertically stratified (i.e. Gaussian) gas density distribution with scale height $H$. The gas density is uniform along $(x,y)$ apart from small fluctuations induced to seed the SI. Upon this relatively smooth background, we generate a Gaussian (centered on $x = 0$, with radial width $w = 1.14H$) in the gas density and pressure. The target density profile of the pressure bump is given by

\begin{equation}
  \label{eqn:rho_bump}
  \hat{\rho}(x,y,z) = \rho_0 \left[ 1 + A e^{\left(-x^2/2w^2\right)}\right] e^{\left(-z^2/2H^2\right)},
\end{equation}

\noindent
where $A$ is the bump amplitude, a free parameter in our simulations, $\rho_0$ is the initial mid-plane gas density, and $H$ is the vertical gas scale height. A pressure bump with that profile is in geostrophic balance when the azimuthal gas velocity is

\begin{equation}
  \label{eqn:uy_bump}
  \hat{u}_y(x,y,z) = \frac{-A x\cs^2e^{\left(-x^2/2w^2\right)}}{ 2w^2\Omega\left[ 1 + A e^{\left(-x^2/2w^2\right)}\right]}
\end{equation}

\noindent
where $\cs$ is the (isothermal) gas sound speed, and $\Omega$ is the orbital frequency at the center of the shearing box. Note that the gas velocity is in a reference frame where the background Keplerian shear has been subtracted (i.e., the shearing frame). The bump is created and maintained using Newtonian relaxation over a timescale $\treinf$, which is the second free parameter in our simulations. At each timestep $\Delta t$ we adjust the radial profile of $\rho$ and $u_y$ by

\begin{eqnarray}
  \label{eqn:Delta_rho}
  \Delta \rho &=& (\hat{\rho} - \rho)  \frac{\Delta t}{\treinf} \\
  \label{eqn:Delta_uy}
  \Delta u_y  &=& (\hat{u}_y  - u_y )  \frac{\Delta t}{\treinf}
\end{eqnarray}

The simulation box is placed at 50 AU within the model described above (and in \citealt{Carrera_2021}) with a pressure bump that is essentially modeled after one of the rings of HL Tau. The initial concentration of solids is set to the solar value:

\begin{equation}
Z \equiv \frac{\Sigma_p}{\Sigmag} = 0.01,
\end{equation}

\noindent 
where $\Sigmag$ and $\Sigmap$ are the gas and particle surface densities. The solid component is implemented as Lagrangian super-particles (see \citealt{Bai_2010a}), initialized as a Gaussian density distribution with scale height $H_p = 0.025 H$. The background dimensionless pressure gradient (not accounting for the pressure bump) is

\begin{equation}
    \label{eqn:PI}
    \Pi \equiv \frac{\Delta v}{\cs}
        = - \frac{1}{2} \left( \frac{\cs}{\vk} \right) \frac{d \ln P}{d \ln r} = 0.12 
\end{equation}
where $\vk$ is the Keplerian velocity, $P$ is the gas pressure, $r$ is the radial distance away from the central star, and $\Delta v$ is the headwind experienced by solid particles

\begin{eqnarray}
    \Delta v &\equiv& \vk - u_\phi = \eta \vk \\
    \eta &=& - \frac{1}{2} \left( \frac{\cs}{\vk} \right)^2 \frac{d \ln P}{d \ln r}.
\end{eqnarray}

\noindent
The strength of self-gravity is defined via the $\tilde{G}$ parameter:

\begin{equation}
    \tilde{G} \equiv \frac{4\pi G \rho_0}{\Omega^2} \approx 0.2,
\end{equation}

\noindent
This corresponds to a Toomre \citep{Toomre_1964} $Q$ value of $Q \approx 8$. 

\noindent
The Stokes number is defined as

\begin{equation}
    \label{eqn:tstop}
    \tau \equiv \frac{\rhos a}{\rho H} \sqrt{\frac{\pi}{8}},
\end{equation}

\noindent
where $a$ is the particle size, $\rho$ is the gas density, and $\rhos$ is the material density of solids. The Stokes number is calculated dynamically throughout the simulation, assuming a constant value of the particle size of $a = 1$~cm for all runs. 

Since our runs are primarily parameterized by $\treinf$ and $A$, our runs all follow a naming convention where, for example, run \texttt{Tr10-A15} has $\treinf = 10\Omegainv$ and $A = 0.15$. Our full list of runs are described in the next section and summarized in Tables \ref{table:runs-1} and \ref{table:runs-2}.

\subsection{Experiment Setup}
\label{sec:methods:experiment}

Our investigation has two main goals:

\begin{enumerate}
\item \textit{Find the conditions for planetesimal formation in the $A$-$\treinf$ parameter space.}

We run a grid of simulations with more granular bump amplitudes $A \in \{ 0.05, 0.10, 0.15, 0.20 \}$ and more reinforcement times $\treinf \in \{1, 10, 100\} \Omegainv$ than in \cite{Carrera_2021}. These runs are listed in Table \ref{table:runs-1}.

\item \textit{Find the effect of $\treinf$ on planetesimal formation.}

We fix the pressure bump amplitude at $A = 0.20$ and for each $\treinf \in \{1, 10, 100\} \Omegainv$ we run six simulations with different (random) initial particle positions and gas velocity perturbations. We measure the time, location, and number of planetesimal formation events, the planetesimal formation efficiency, and the pebble accretion rate. These runs are listed in Table \ref{table:runs-2}.
\end{enumerate}

\begin{figure}[ht]
    \centering
    \includegraphics[width=1\linewidth]{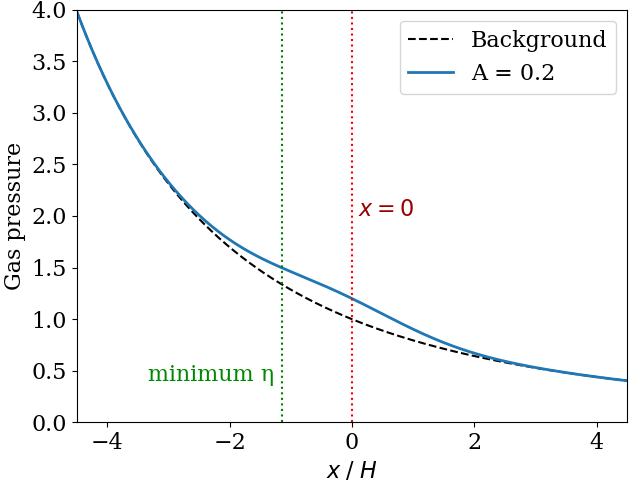}
    \caption{Gas pressure profile with a pressure bump (blue, $A = 0.2$) and without (dashed black). Because the pressure bump is sitting on top of a background pressure gradient, the point of minimum headwind is offset from the center of the simulation box.}
    \label{fig:density_profile}
\end{figure}

Because we run many more simulations than \citet{Carrera_2021}, we need to run our simulations at slightly lower resolution (320/H vs 640/H). While we use the same width and length ($L_x \times L_y = 9H\times 0.2H$), we use taller boxes. The appropriate box height $L_z$ was determined from a convergence test. Finally, Figure \ref{fig:density_profile} shows the net pressure density profile of our simulated disk for a typical bump amplitude of $A = 0.2$. Notice that the point of minimum headwind, where the SI should be most effective, is offset from the center of the box. Notice also that for $A = 0.2$ there is no local pressure maximum, and therefore no particle trap.

\begin{table}[ht]
    \caption{We ran a grid of simulations with varying pressure bump amplitude $A$ and reinforcement time $\treinf$. Goal 1 of this work is to probe the region of the $A$ vs $\treinf$ parameter space that is consistent with planetesimal formation.}
   \label{table:runs-1}
    \centering
    \begin{tabular}{lcr}
      Run   &  Amplitude  & $\treinf$ \\
	\hline
    \texttt{Tr1-A05}   & 0.05 & $1\Omegainv$  \\
    \texttt{Tr1-A10}   & 0.10 & $1\Omegainv$  \\
    \texttt{Tr1-A15}   & 0.15 & $1\Omegainv$  \\
    \texttt{Tr1-A20}   & 0.20 & $1\Omegainv$  \\
    \hline
    \texttt{Tr10-A15}  & 0.15 & $10\Omegainv$  \\
    \texttt{Tr10-A20}  & 0.20 & $10\Omegainv$  \\
    \hline
    \texttt{Tr100-A15} & 0.15 & $100\Omegainv$  \\
    \texttt{Tr100-A20} & 0.20 & $100\Omegainv$  \\
    \end{tabular}
\end{table}

\begin{table}[ht]
    \caption{With a fixed pressure bump amplitude $A = 0.2$ we ran six simulations for each reinforcement time $\treinf \in \{1, 10, 100\}/ \Omega$; each simulation was initiated with a different random seed. Goal 2 of this work is to determine the effect of $\treinf$ on the time, location, and properties of planetesimal formation events.}
    \label{table:runs-2}
    \centering
    \begin{tabular}{lccr}
      Run   &  $L_z$ & Amplitude & $\treinf$ \\
    \hline
    \texttt{Tr1-Seed[1..6]}    & $6.4 H$  & 0.2  & $1\Omegainv$ \\
    \texttt{Tr10-Seed[1..6]}   & $6.4 H$  & 0.2  & $10\Omegainv$ \\
    \texttt{Tr100-Seed[1..6]}  & $6.4 H$  & 0.2  & $100\Omegainv$ \\
    \end{tabular}
\end{table}

%
%
\section{Results}
\label{sec:results}

\subsection{Is a Bump Needed at All?}
\label{sec:results:bump_needed}

For our disk model, our $a = 1$cm particles correspond to a Stokes number of $\tau \approx 0.12$. When comparing this value to past estimates of the critical $Z$ needed to form particle filaments \citep[e.g.,][]{Li_2021} we must recall that those results are conditioned on a particular choice of $\Pi$. \citet{Sekiya_2018} showed that normalizing $Z$ by the pressure gradient $\sigma \equiv Z/\Pi$ gives a better predictor of when particle clumps will form. The limits of \citet{Li_2021} were done with $\Pi = 0.05$. Extrapolating to $\Pi \approx 0.10$ for a small pressure bump in our disk model, one can expect that $Z > 0.01 $ may be needed to form filaments. Furthermore, even if filaments form, it does not follow that planetesimals will form as well.

\subsection{Convergence Test}
\label{sec:results:convergence}

Before our main investigation, we run a convergence test. Previous authors have found that particle feedback may distort the shape of a gas pressure bump \citep[][]{Taki_2016,Carrera_2021}. Therefore, one must ensure that any bump distortion is physical. For example, most SI simulations have a small domain of $L_z \le 0.4H$, which contains only $\le 16\%$ of the total gas mass. Hence, there may be interactions between the gas and solids that are not captured within such a small domain.

\begin{figure}[ht]
    \centering
    \includegraphics[width=1\linewidth]{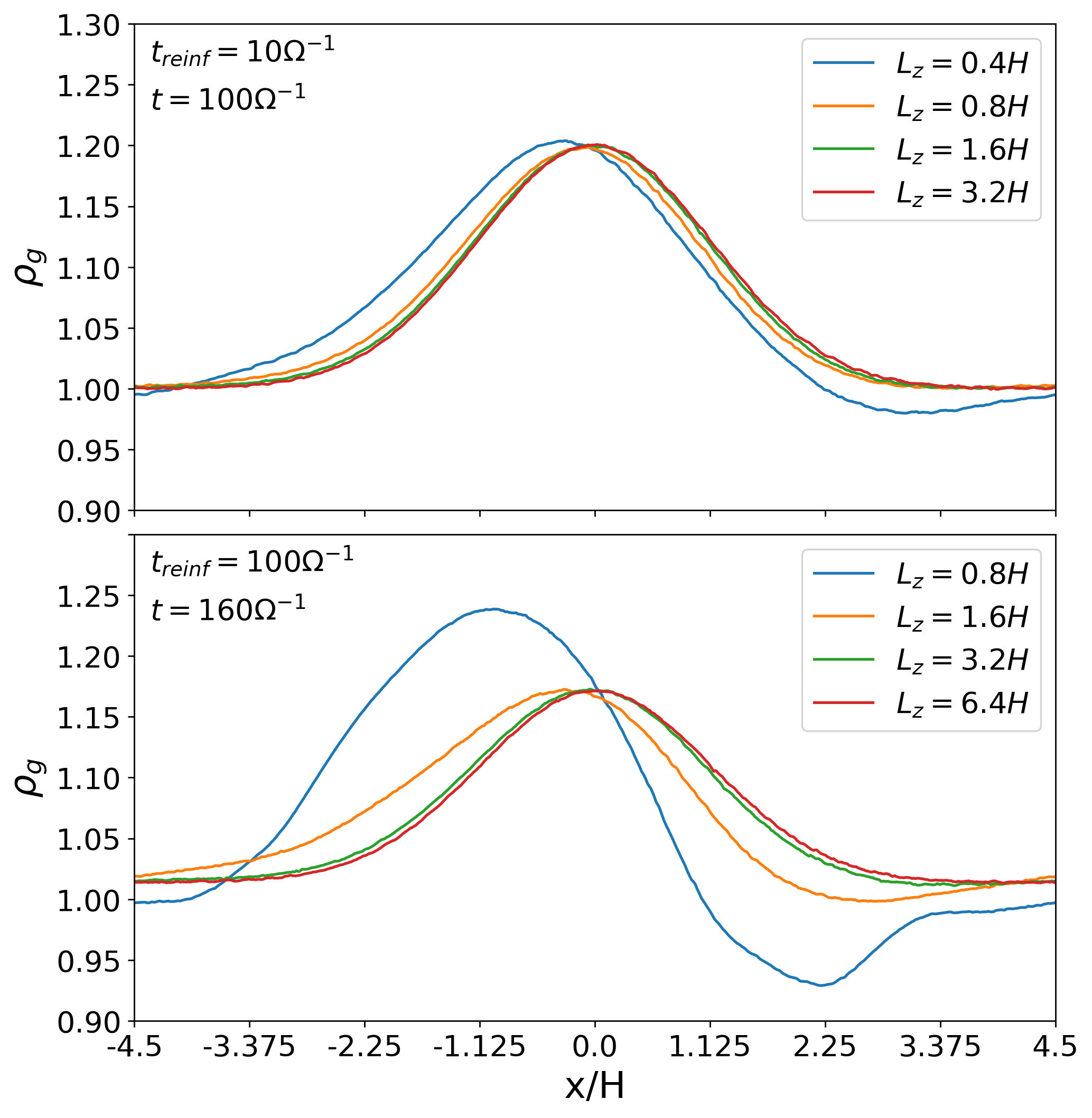}
    \caption{Convergence test for simulations with weakly reinforced bumps ($\treinf \ge 10\Omegainv$). The plots show the midplane gas density (in code units) for increasingly tall simulation boxes. As $L_z$ increases, the amount of gas inside the box grows and the density profile converges toward the targeted (Gaussian) profile (Equation \ref{eqn:rho_bump}).}
    \label{fig:convergence}
\end{figure}

Figure \ref{fig:convergence} shows the midplane gas density profile for simulations with $\treinf \in \{ 10, 100 \} \Omegainv$ and progressively taller simulation domains. For $L_z = 0.4H$ the distortion of the pressure bump is quite significant, but as $L_z$ increases (so that the simulation contains more gas) the pressure bump settles toward the Gaussian profile targeted by our Newtonian relaxation scheme.

\begin{itemize}
\item For $\treinf = 10\Omegainv$, runs with $L_z = 1.6H$ and $3.2H$ give nearly identical gas profiles (they differ by $\le 0.04\%$ everywhere). Therefore it appears that the $L_z = 1.6H$ run is fully converged.

\item For $\treinf = 100\Omegainv$, only the run with $L_z = 6.4H$ appears to be converged. At that box size, the simulation already contains almost 99.9\% of the gas mass and the bump profile differs from the idealized Gaussian by $\le 0.02\%$ everywhere.
\end{itemize}

Therefore, we find that $L_z = 1.6H$ and $6.4H$ are the minimum box sizes needed to model $\treinf = 10\Omegainv$ and $100\Omegainv$ respectively.

We are not entirely sure why the solution converges with increasing box size.  However, we speculate that it is related to the fact that particles only come in contact with the gas at the midplane. A very thin box would be heavily particle-dominated, and as such the dynamics of the gas at the mid-plane would be largely controlled by the particles with which it is interacting. However, a tall box has sufficient gas that despite the distortion of the gas in the mid-plane, the gas {\it away} from the mid-plane controls the dynamics.  In the simplest terms, the gas at the mid-plane is given angular momentum by the particles (and thus moves outwards), yet the gas above and below the mid-plane has not been distorted and thus readily fills any deficit of gas in the mid-plane.  This is only an issue as $\treinf$ gets larger because for short reinforcement times, the mid-plane gas is adjusted back to its original position ``by hand".

\subsection{Conditions for Planetesimal Formation}
\label{sec:results:conditions}

\begin{figure}[ht]
    \centering
    \includegraphics[width=1\linewidth]{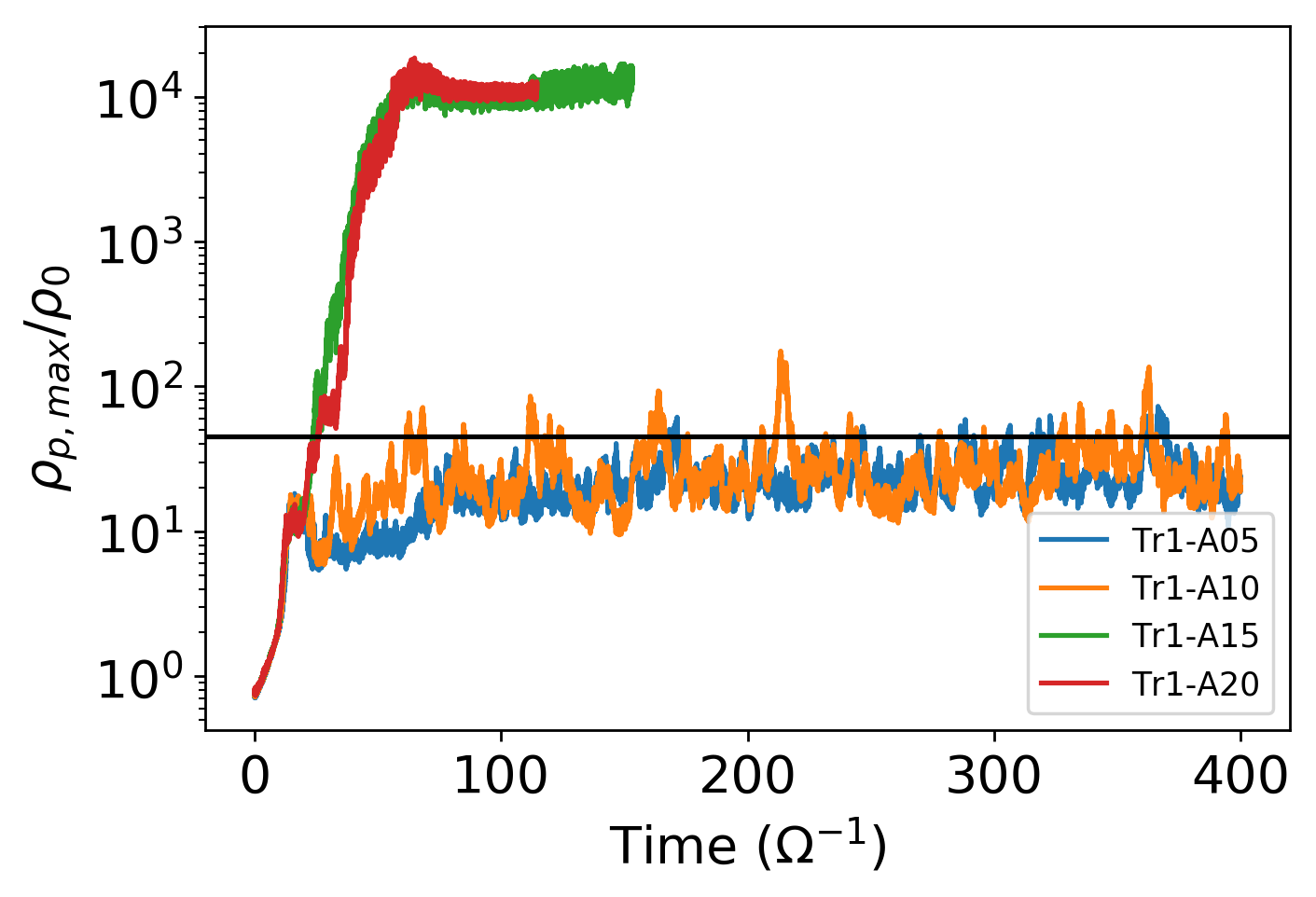}
    \caption{Maximum particle density for four simulations with $\treinf = 1\Omegainv$ and $0.05 \le A \le 0.20$. Here $\rho_0$ is the gas density away from the bump and the solid line marks the Roche density. The initial sedimentation takes $t \approx 10 \Omegainv$. After that, the SI forms filaments, and if the filaments cross the Roche density, gravitational collapse follows. The simulations clearly form gravitationally bound particle clumps for $A \ge 0.15$. For $A = 0.10$ there is a very small bound clump at $t \approx 410\Omegainv$ but it appears to be a fluke (see main text).
    }
    \label{fig:dmax_A}
\end{figure}

Figure \ref{fig:dmax_A} shows the maximum particle density for four simulations with $\treinf = 1\Omegainv$ and $0.05 \le A \le 0.20$. The box size is $L_x \times L_y \times L_z = 9H\times 0.2H\times0.4H$. The density is normalized by the midplane gas density away from the bump $\rho_0$; on this scale, the Roche density is $\rho_{\rm roche} \approx 45 \rho_0$.

We find that dense particle clumps form quickly for bumps with amplitude $A \ge 0.15$, which is a smaller value than the limit found by \citet{Carrera_2021}. For $A = 0.10$ the simulation forms a single clump at $t \approx 410\Omegainv$. However, this is an extremely small clump containing only 0.3\% of the total number of particles, while runs with $A \ge 0.15$ have $\sim 10\%$ of particles in bound clumps. Table \ref{table:mass} shows the fraction of particles incorporated into clumps for $A \ge 0.10$. Therefore, we conclude that the one small clump in $A = 0.10$ was a fluke and treat the run as a failure to produce planetesimals.

\begin{table}[ht]
    \caption{Fraction $f$ of particles incorporated into bound clumps; defined as as particles in grid cells with density above $\rho_{\rm roche}$ or $10\rho_{\rm roche}$ at the end of each simulation.}
    \label{table:mass}
    \centering
    \begin{tabular}{crr}
      Run   &  $f[\rhop > \rho_{\rm roche}]$ & $f[\rhop > 10\rho_{\rm roche}]$ \\
	\hline
    \texttt{Tr1-A20}  & 10.7\%  &   9.8\% \\
    \texttt{Tr1-A15}  & 11.2\%  &  10.0\% \\
    \texttt{Tr1-A10}  &  0.3\%  &   0.2\% \\
    \end{tabular}
\end{table}

\begin{figure*}[ht!]
    \centering
    \includegraphics[width=0.75\textwidth]{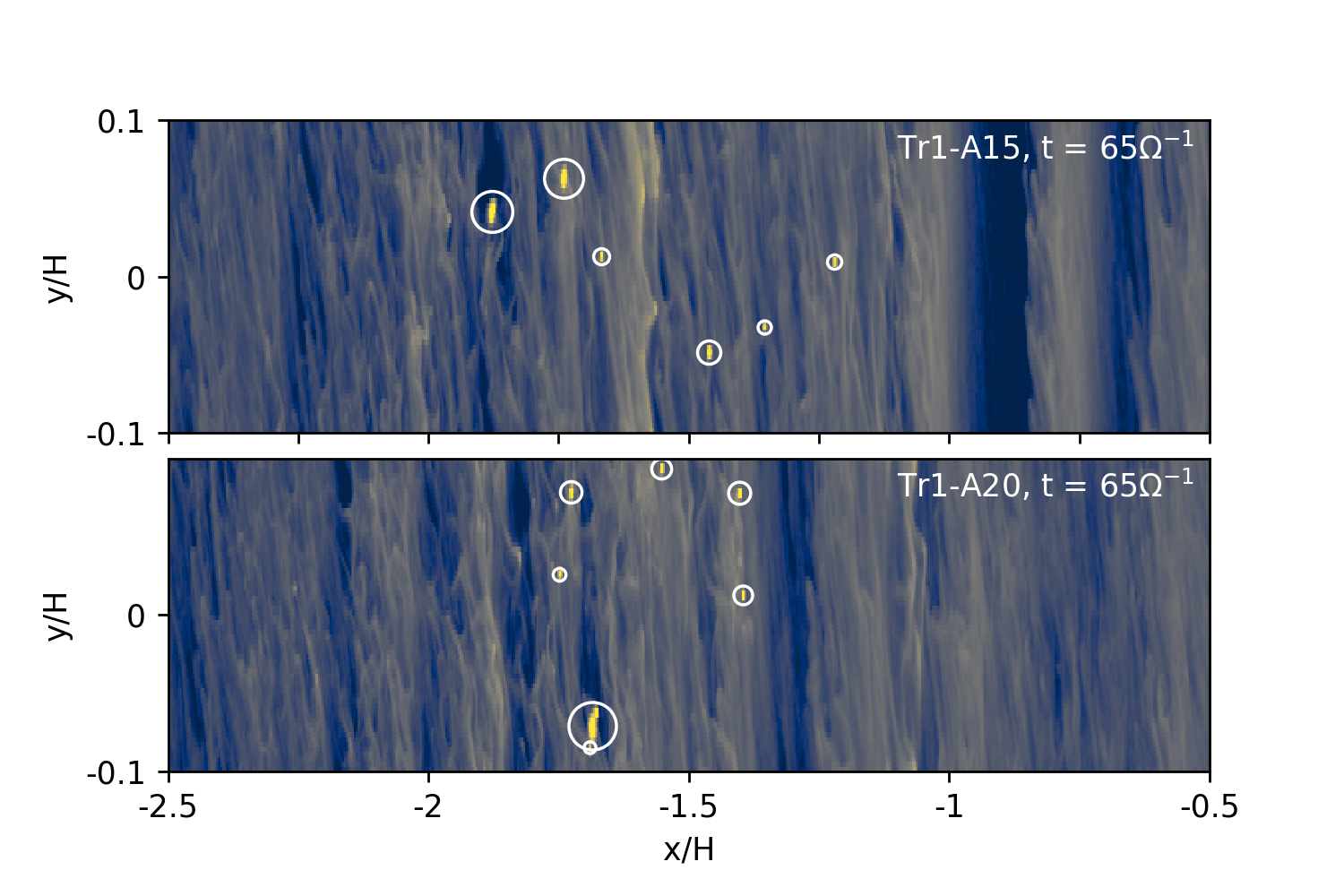}
    \raisebox{0.14\height}{\includegraphics[width=0.11\textwidth]{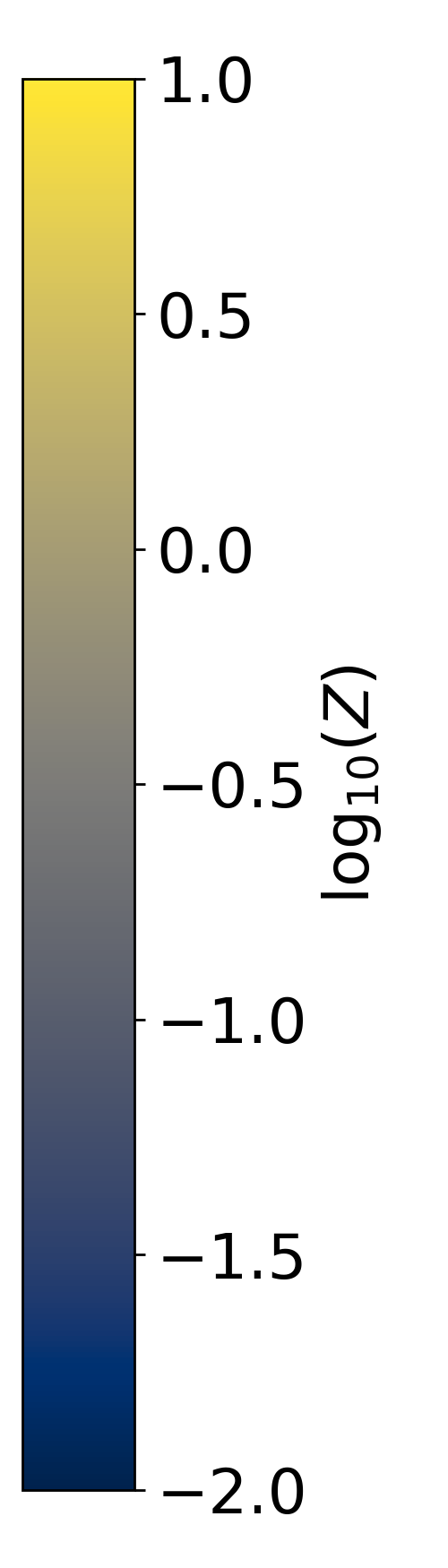}}
    \caption{Snapshots of the dust-to-gas ratio $Z = \Sigmap / \Sigmag$ for $\treinf = 1\Omegainv$ and $A \in \{0.15, 0.20 \}$ (i.e., models \texttt{Tr1-A15} and \texttt{Tr1-A20}) near the end of the simulation. The white circles mark the particle clumps with densities $\rhop > 1000\rhog$. The clumps are roughly spherical, but the aspect ratio of our plot makes them appear elongated.}
    \label{fig:2D_plot}
\end{figure*}

Figure \ref{fig:2D_plot} shows snapshots of the dust-to-gas ratio for the two runs with $A \ge 0.15$. In addition to gravitationally bound particle clumps, the filaments associated with the SI are clearly visible. The clumps form at roughly the same location and in similar quantities --- there are generally one or two very massive clumps, and a handful of smaller ones. The densest clumps, with $\rhop > 1000\rhog$, are marked with white circles, and the diameter of each circle is proportional to $M_{\rm clump}^{1/3}$. The clumps span an order of magnitude in mass from $\sim 0.5\%$ to $\sim 5\%$ of the total particle mass. These values were measured at the time when clumps first form ($t = 65\Omegainv$) to avoid the effect of pebble accretion, which we cannot simulate accurately with the gravitational potential discretized on a finite grid.


Next, we ran four additional simulations with $\treinf \in \{10, 100\}\Omegainv$ and $A \in \{0.15, 0.20\}$. For these runs we increased the height of the box to $L_z = 1.6H$ and $6.4H$ respectively to ensure convergence. Those runs all produced planetesimals. We did not conduct any simulations with $\{ \treinf \ge 10\Omegainv, A \le 0.10\}$ because we reasoned that if planetesimals cannot form for $\treinf = 1\Omegainv$ then they will not form for $\treinf > 1\Omegainv$ either. These results are summarized in Table \ref{table:A_vs_treinf}. This result is our first indication that planetesimal formation is largely insensitive to $\treinf$.

\begin{table}[ht]
    \caption{Table depicting whether or not planetesimals form for different combinations of $A$ and $\treinf$. The ability to form planetesimals does not seem to depend on $\treinf$.}
    \label{table:A_vs_treinf}
    \centering
    \begin{tabular}{c|ccc}
    $\treinf =$
        & $1\Omegainv$
        & $10\Omegainv$
        & $100\Omegainv$ \\
	\hline
	$A = 0.20$ & \cmark & \cmark & \cmark \\
	$A = 0.15$ & \cmark & \cmark & \cmark \\
	$A = 0.10$ & \xmark & \xmark & \xmark \\
	$A = 0.05$ & \xmark & \xmark & \xmark \\
    \end{tabular}
\end{table}

\subsection{Effect of $\treinf$ on Planetesimal Formation}
\label{sec:results:effect}

For our second experiment we ran six simulations for each $\treinf \in \{1, 10, 100\}\Omegainv$ for a total of eighteen runs. Each run had a different random seed to initialize the particle positions and random velocity perturbations of the gas. All runs had a bump amplitude of $A = 0.2$ and to keep the runs as similar as possible they all had the same box height of $L_z = 6.4H$, regardless of $\treinf$. Doing six simulations for each $\treinf$ improves the statistics, which is particularly important here because our relatively low resolution \citep[e.g., versus][]{Simon_2017} results in fewer planetesimals. A similar approach was carried out by \cite{Rucska_2021}.

\begin{figure}[ht]
    \centering
    \includegraphics[width = \linewidth]{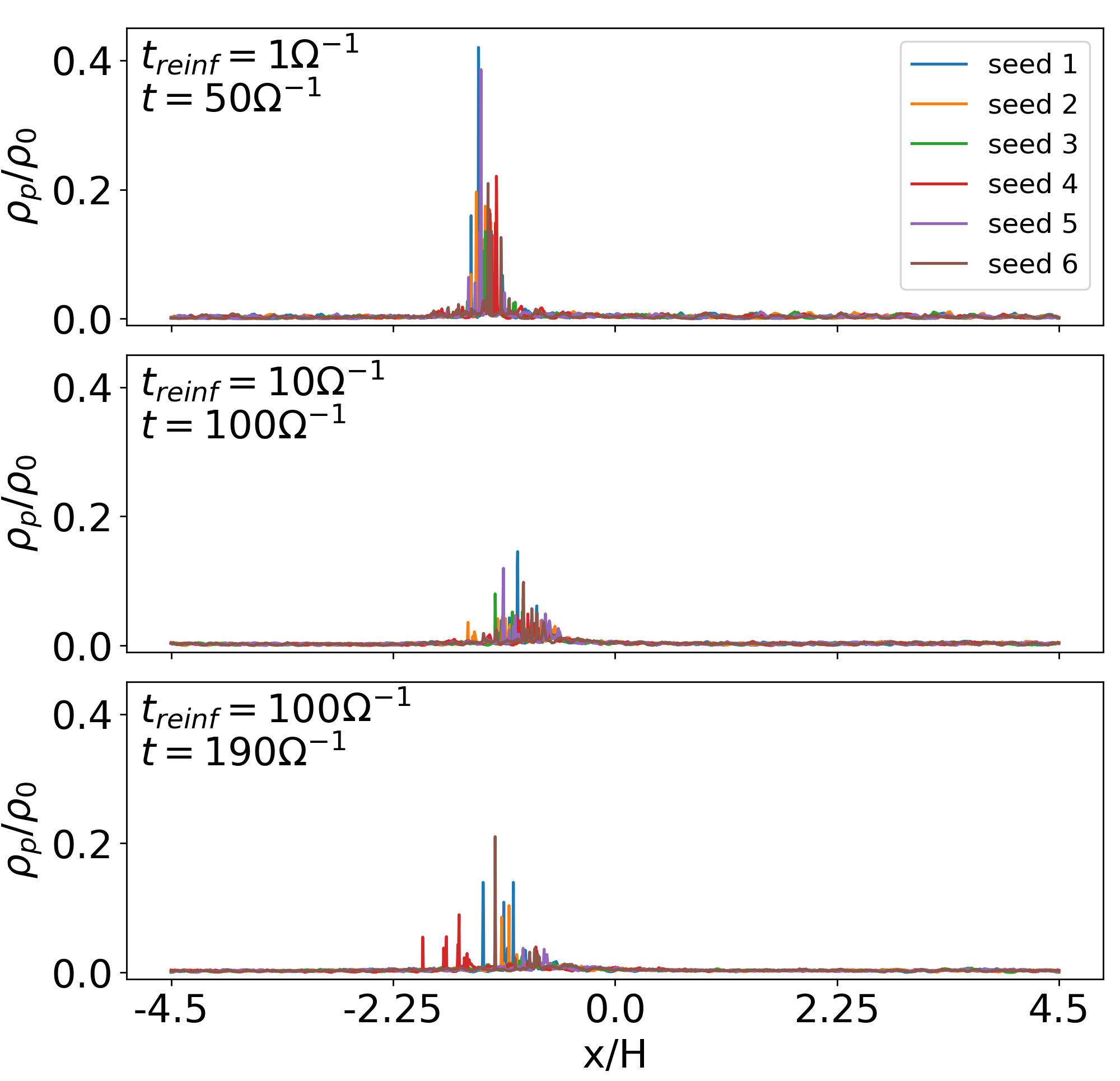}
    \caption{Mean particle density (averaged along $y$ and $z$) normalized by the midplane gas density away from the bump $\rho_0$ for all runs in Table \ref{table:runs-2}. The main take-away is that planetesimal formation events occur at roughly the same location regardless of reinforcement time $\treinf$.}
    \label{fig:Tr1ClumpLocs}
\end{figure}

Figure \ref{fig:Tr1ClumpLocs} shows the locations where bound particle clumps begin to form for all runs. The figure shows that all planetesimal formation events occur at roughly the same location, at around $x \approx -1.2H$. The reasons for that location were discussed in detail by \citet{Carrera_2021}, but the core idea is that planetesimals form near the region of minimum headwind $\eta$. For a pressure bump with amplitude $A = 0.2$ there is no pressure peak, as there is no pressure trap. However, there is a point where $\eta$ is smallest, and planetesimals form near that point.

For $\treinf = 100\Omegainv$ planetesimals form slightly closer to the star, but this may just reflect that at the time the snapshot is taken the $\treinf = 100\Omegainv$ runs have only gone through a couple of e-folding times; so the bump has not yet reached full height. This would be a continuation of the trend seen by \citet{Carrera_2021} where smaller bumps had their planetesimal formation events somewhat closer to the star. The height of the density peaks in Figure \ref{fig:Tr1ClumpLocs} is not significant as it depends on how long the clump has been allowed to accrete particles. We will discuss clump masses and accretion in depth at the end of this section.

\begin{figure}[ht]
    \centering
    \includegraphics[width=\linewidth]{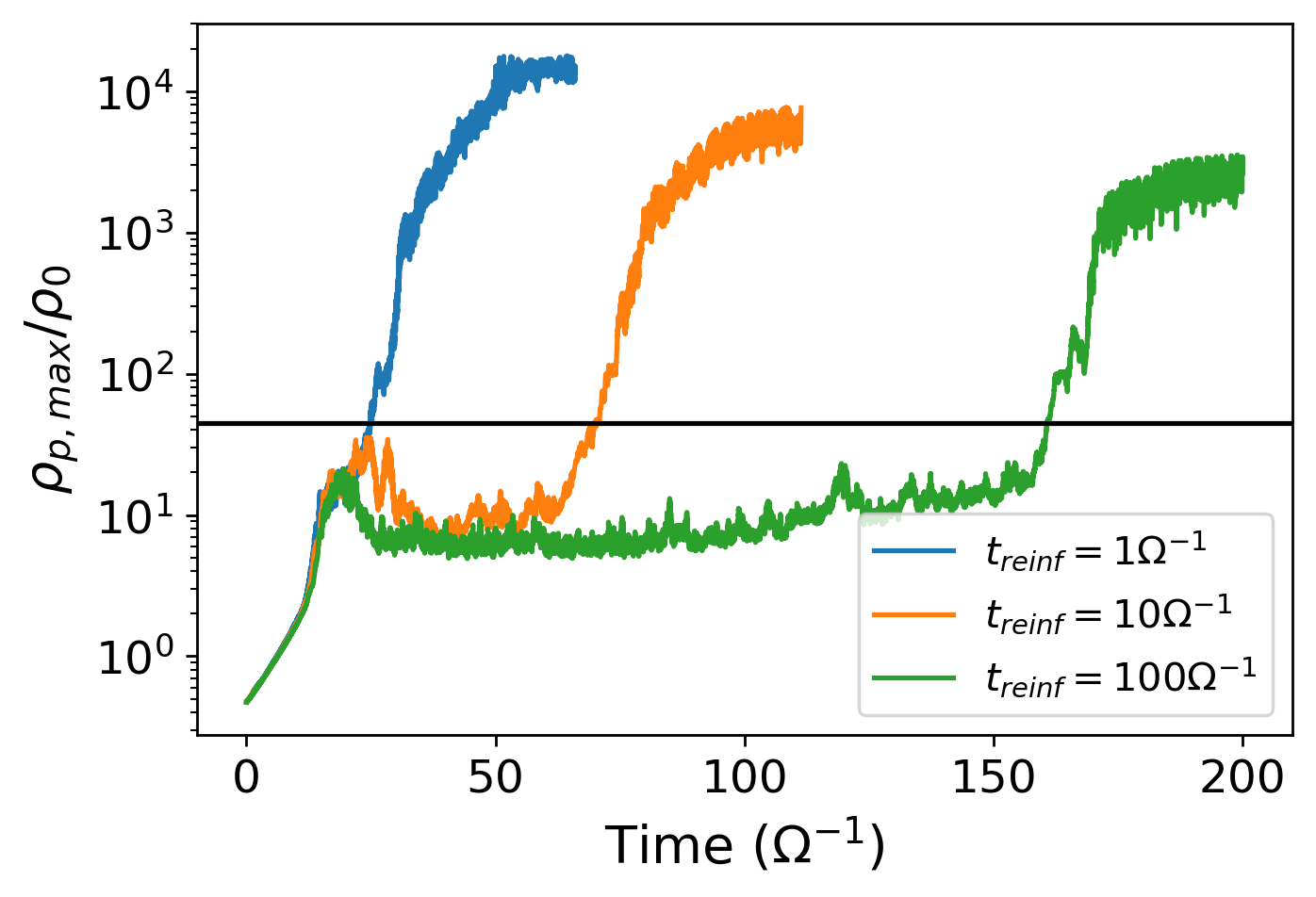}
    \caption{Maximum particle density for three example runs with $\treinf \ge 1\Omegainv$. All runs have the same pressure bump amplitude $A = 0.2$ and the density is normalized by the midplane gas density away from the bump $\rho_0$. The solid line marks the Roche density. The initial sedimentation takes $t \approx 10 \Omegainv$. The delay in planetesimal formation is expected as $\treinf$ is the timescale for the bump to form.}
    \label{fig:dmax_TR}
\end{figure}

\begin{figure}[ht]
    \centering
    \includegraphics[width=\linewidth]{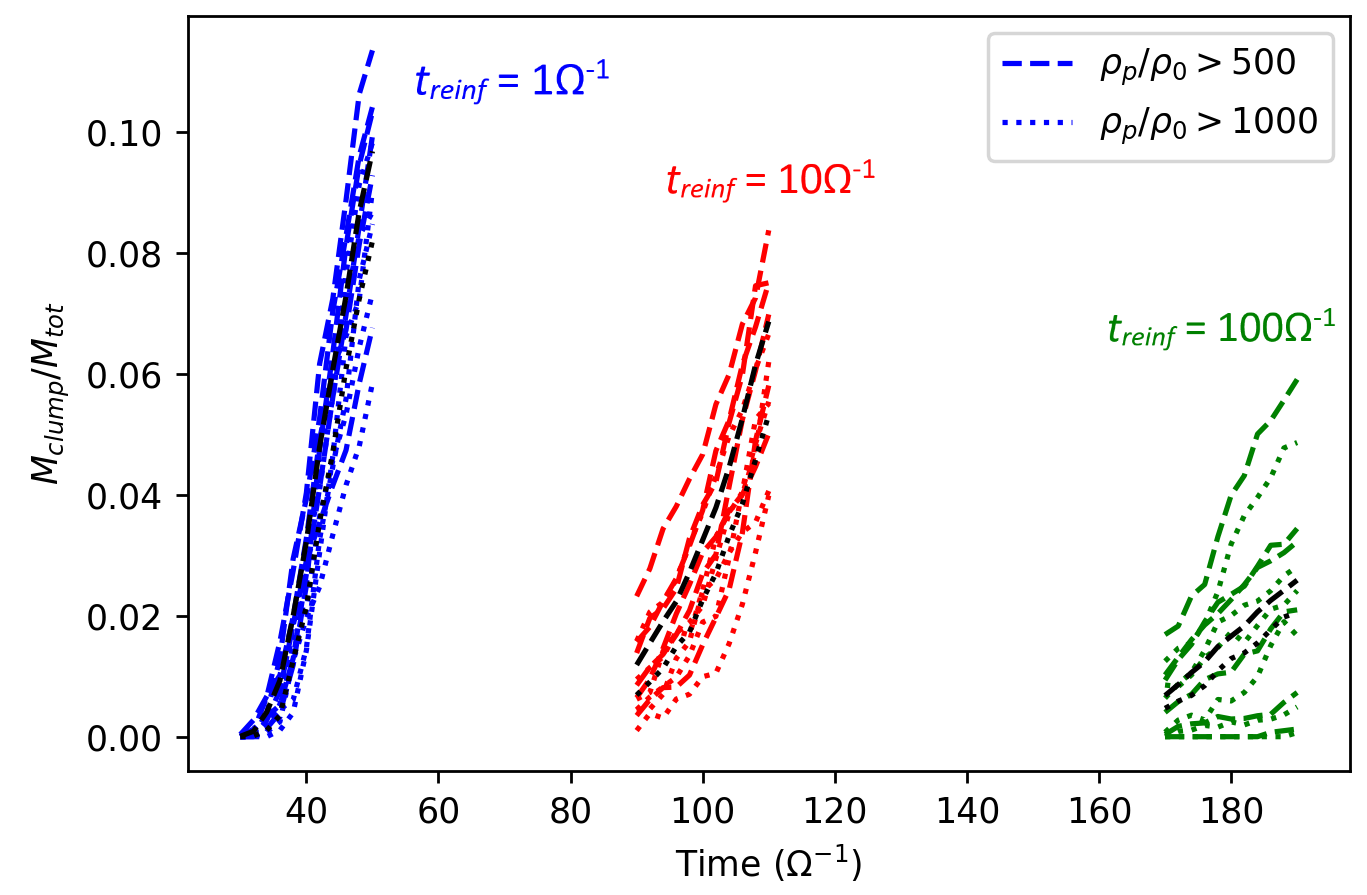}
    \caption{Mass of bound clumps as a function of time for all runs in Table \ref{table:runs-2}. Here $M_{\rm clump}$ is the total particle mass inside grid cells with $\rhop/\rho_0 > 500$ (dashed) or $1,000$ (dotted) where $\rho_0$ is the midplane gas density away from the bump. $M_{tot}$ is the total particle mass in the domain. Lower $\treinf$ consistently leads to clump growth via faster accretion.
    }
    \label{fig:FracMass}
\end{figure}

\begin{figure}[ht]
    \centering
    \includegraphics[width=\linewidth]{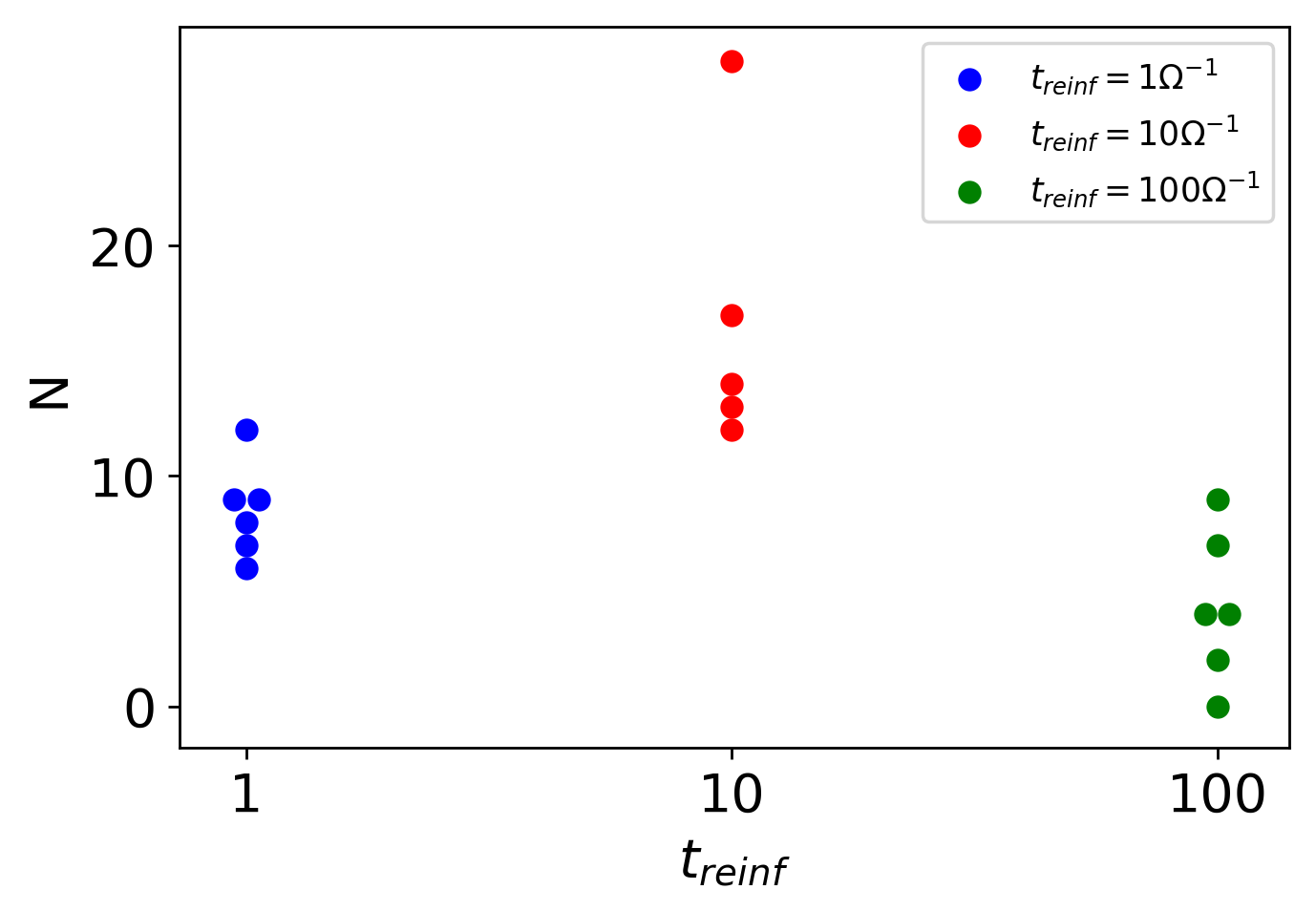}
    \caption{Number of bound particle clumps for all runs in Table \ref{table:runs-2}. There is no discernible trend between reinforcement time $\treinf$ and the number of clumps produced. Note that one of the simulations with $\treinf = 100\Omegainv$ did not form any planetesimals.
    }
    \label{fig:NClumps}
\end{figure}

Figure \ref{fig:dmax_TR} shows the maximum particle density for three example runs with $\treinf \in \{1, 10, 100\}\Omegainv$. All runs formed planetesimals. Unsurprisingly, simulations that form the pressure bump later (i.e., longer $\treinf$) also form planetesimals later. But we \textit{were} surprised to see a trend between $\treinf$ and the mass and density of the clumps (this trend holds across runs) --- clumps from strongly reinforced bumps (small $\treinf$) are denser and more massive. Figure \ref{fig:FracMass} shows the total particle mass inside clumps $M_{\rm clump}$ as a function of time for all runs in this experiment. For the purpose of that figure we define $M_{\rm clump}$ as the total particle mass inside grid cells with $\rhop/\rho_0 > 500$ or $1,000$ (both are shown). We emphasize that the growth rate of $M_{\rm clump}$ is not due to new clumps forming, but to existing clumps accreting faster when $\treinf$ is smaller. More specifically, we used the Planetesimal Analyzer ({\sc PLAN}) tool of \citet{Li_2019} to quantify the number of clumps at the end of each run and found (see Fig.~\ref{fig:NClumps}) that the number of clumps (around $\sim 10$) is not correlated with $\treinf$. Therefore, the differential growth rates of $M_{\rm clump}$ in Figure \ref{fig:FracMass} reflect a difference in pebble accretion rates.

\subsubsection{Pebble Accretion Rate}
\label{sec:results:effect:accretion}

The pebble accretion rate is over-estimated in our simulations. Like other works on the SI, we have a finite grid resolution with grid cells at scales {\it much} larger than typical planetesimal sizes; therefore we cannot resolve the final collapse of the particle clump into planetesimals. Real planetesimals would have a much smaller accretion cross-section than our clumps. Still, we must determine why the simulation shows a correlation between $\treinf$ and accretion rate and whether a similar process may occur for real planetesimals.

First, we shall focus on runs with $\treinf < 100\Omegainv$ because our $\treinf = 100\Omegainv$ do not run long enough for their pressure bumps to reach their full height. Comparing $\treinf = 1\Omegainv$ versus $10\Omegainv$, we found that clumps in low-$\treinf$ runs accrete faster \textit{because pebbles drift faster}, and therefore the pebble flux is higher. The mean slope in Figure \ref{fig:FracMass} ($\approx$ accretion rate) is around $2.0 \pm 0.2$ higher for $\treinf = 1\Omegainv$ than for $\treinf = 10\Omegainv$. Over the same time periods, the particle drift rate for $\treinf = 1\Omegainv$ is $2.19 \pm 0.03$ times faster than for $\treinf = 10\Omegainv$. 



%
%
\section{Discussion}
\label{sec:discussion}

\subsection{Particle Drift}
\label{sec:discussion:drift}

\textit{Why is $\treinf$ linked to particle drift?} Any force that maintains a pressure bump (e.g., planets, MHD winds, zonal flows) must do so by removing angular momentum from the system: Aerodynamic drag causes pebbles to give their angular momentum to the gas (for $dP/dr < 0$, as in our runs); that angular momentum must be removed from the gas to reinstate the bump (thus angular momentum is not conserved in our runs). This negative torque is transferred to the particles effectively immediately (since $\tau \ll \treinf$). Eventually the system will reach an equilibrium somewhat analogous to the NSH equilibrium of \citet{Nakagawa_1986}, but with additional terms. In the absence of external forces, the equations of motion in the co-rotating and shearing frame are

\begin{eqnarray}
\label{eqn:NSH:Vx}
\frac{\partial V_x}{\partial t}
    &=&
    -C\rho(V_x - u_x) + 2\Omega V_y\\
\label{eqn:NSH:Vy}
\frac{\partial V_y}{\partial t}
    &=&
    -C\rho(V_y - u_y) - \frac{1}{2}\Omega V_x\\
\label{eqn:NSH:ux}
\frac{\partial u_x}{\partial t}
    &=&
    -C\rhop(u_x - V_x) + 2\Omega u_y
    -\frac{1}{\rho}\frac{\partial P}{\partial x}\\
\label{eqn:NSH:uy}
\frac{\partial u_y}{\partial t}
    &=&
    -C\rhop(u_y - V_y) - \frac{1}{2}\Omega u_x
\end{eqnarray}

\noindent
where $(u_x, u_y)$ and $(V_x,V_y)$ are the velocities of the gas and particle components respectively, and $C$ is the drag coefficient of the particle components. \citet{Nakagawa_1986} showed that in a steady state (i.e., $\partial / \partial t = 0$) the radial velocity of the particle component is

\begin{equation}
    \left.V_x\right|_{\rm NSH}
    =
    - \frac{\rho}{\rho + \rhop}
    \frac{2D\Omega}{D^2 + \Omega^2} \eta\vk
\end{equation}

\noindent
where $D = A(\rho + \rhop)$.

In the case of a disk with a pressure bump, there are additional force terms in the gas equations (Equations \ref{eqn:NSH:ux} and \ref{eqn:NSH:uy}), and those terms will ultimately be reflected in the solution for $V_x$. In the case of our Newtonian relaxation scheme, there is a $\Delta \rho$ component (Equation \ref{eqn:Delta_rho}) that alters the gas density profile which sets the pressure gradient term $(1/\rho)dP/dx$, and there is a $\Delta u_y$ component (Equation \ref{eqn:Delta_uy}) that adds a force term to the equation for $\partial u_y/\partial t$

\begin{equation}
    \frac{\partial u_y}{\partial t}
    =
    -A\rhop(u_y - V_y) - \frac{1}{2}\Omega u_x
    -\frac{u_y - \hat{u}_y}{\treinf}
\end{equation}

For conciseness, let $t_r \equiv \treinf$, $\epsilon = \rhop/\rho$. The steady state solution is

\begin{eqnarray*}
V_y &=& 
    - \left(
        \frac{D^2 + \Omega^2}{A\rho} +
        \frac{\epsilon}{t_r}
    \right)^{-1}
    \left(
        (D + \frac{1}{t_r}) \eta \vk +
        \frac{\hat{u}_y}{t_r}
    \right) \\
u_y &=&
    - \eta \vk - \epsilon V_y \\
V_x &=&
    2\frac{\hat{u}_y}{t_r\Omega}
    + 2\left(
        \frac{\Omega}{A\rho} + \frac{A\rhop}{\Omega}
    \right)  V_y
    - 2 \left(
        \frac{A\rhop}{\Omega} + \frac{1}{t_r\Omega}
    \right) u_y
\end{eqnarray*}

\noindent
As $t_r \rightarrow \infty$ we recover the solution of \citet{Nakagawa_1986}. As $t_r \rightarrow 0$,

\begin{eqnarray*}
V_y &\rightarrow&
    - \frac{\eta\vk + \hat{u}_y}{\epsilon}\\
u_y &\rightarrow& \hat{u}_y\\
V_x &\rightarrow& 
    2\left(
        \frac{\Omega}{A\rho} + \frac{A\rhop}{\Omega}
    \right)  V_y
    - 2 \frac{A\rhop}{\Omega} \hat{u}_y
\end{eqnarray*}
The system converges to a steady state with a finite particle drift rate and non-zero torque. Critically, we find that for $t_r > 0$, $u_y \ne \hat{u}_y$ and therefore the torque is never zero. Furthermore, we have verified numerically that particles drift faster for smaller $t_r$.

To add just one more layer of complexity, only the gas near the midplane is in contact with the particles. For the layer above the dust layer (e.g., $0.01 \lesssim |z/H| \lesssim 0.1$) the 1D solution is simply $u_y = -\eta \vk$, so that there is some vertical velocity shear. Since the gas on that layer is also reinforced, and turbulence puts the layers in contact, it may be that the gas away from the midplane is an important angular momentum reservoir that affects the dust drift velocity, and therefore the pebble accretion rate.

While it is clear that a process that forms and sustains a pressure bump must do so by exchanging angular momentum with the gas, and some of that torque will go to the solid component, additional work is needed to determine whether the link to pebble flux is a real phenomenon or an artifact of our simulations. For instance, a planet producing a pressure bump would necessarily gain the angular momentum lost by the gas; it is not clear how this exchange of angular momentum would change the pebble drift rate result we have found here. To investigate this hypothesis will require new simulations with a more physical source for the pressure bump.

\subsection{Turbulence}
\label{sec:discussion:turbulence}

One limitation of this work is that it does not include an external source of turbulence; our runs only have the small level of turbulence induced by the SI itself. However, the observed ring-like structures probably require a fair amount of radial diffusion \citep[e.g.,][]{Dullemond_2018a}. The role of turbulence on planetesimal formation is not well understood. \citet{Gole_2020} ran simulations with forced turbulence imposing a Kolmogorov energy spectrum and found that that type of turbulence interfered with planetesimal formation. However, \citet{Yang_2018} ran ideal and non-ideal MHD simulations and found strong clumping in an Ohmnic dead zone when solid abundances are similar to
the critical value for a laminar environment. While their ideal-MHD runs produced a mostly isotropic velocity dispersion, their non-ideal runs had lower radial and azimuthal diffusion which facilitated particle clumping. Furthermore, even in a dead zone, the upper layers ($|z| \gtrsim H$) are quite diffusive and isotropic and that might be enough to produce the observed ring-like structures.

In any case, we have opted to ignore turbulence in these runs for simplicity.  We will include its effect in a future study.

\subsection{Why are Dust Rings Still Visible}
\label{sec:discussion:why_visible}

We have shown that planetesimal formation by the SI remains efficient for cm-size particles even for weakly reinforced pressure bumps. This begs the question, if planetesimal formation is fast inside dust rings, why do we see dust rings at all? The most natural explanation is that planetesimal formation removes only a portion of the solids, and preferentially removes larger particles, until the SI stops being efficient. At some point the dust ring should reach a new steady state with a lower planetesimal formation efficiency, where the planetesimal production rate matches the rate of dust drift into the pressure bump.

Looking into this gradual convergence into a steady state is beyond the scope of of the current paper. However, this idea was recently explored by \citet{Stammler_2019}. They ran a 1D disk model with dust coagulation, fragmentation, and drift, as well as a simple prescription for planetesimal formation by the SI. They found that this model explained the apparently fine-tuned optical depth in dust rings studied in the DSHARP survey \citep{Dullemond_2018a}.

%
%
\section{Summary and Conclusions}
\label{sec:conclusions}

We conducted computer simulations to investigate how SI-induced planetesimal formation operates in gas pressure bumps with different amplitudes and reinforcement timescales. Our main results are as follows.

\begin{enumerate}
\item Planetesimal formation is very resilient to weakly-reinforced pressure bumps (large $\treinf$). Planetesimals form for pressure bumps with small amplitudes ($A \ge 0.15$) independently of $\treinf$ (for $\treinf$ as large as $100 \Omegainv$).

\item For the values of $\treinf$ explored here, reinforcement strength has little effect on the number of planetesimals formed, their initial masses, or their location. 

\item Strong reinforcement (small $\treinf$) is associated with much faster growth in planetesimal mass.  This growth rate is caused by a faster particle drift for small $\treinf$, which results in faster pebble accretion. We propose that stronger reinforcement extracts angular momentum more quickly from the gas to reinstate the bump and some of that torque is rapidly transmitted to the solid component.
\end{enumerate}

The robustness of planetesimal formation to the properties of pressure bumps suggest that any number of bump-forming mechanisms could be responsible for inducing this important stage in planet formation.  Indeed, while ALMA has shown us that such pressure bumps are a ubiquitous feature of protoplanetary disks, the mechanisms to form these bumps remain largely unconstrained. 

Within the context of prior theoretical work, our results suggest that the SI criteria often used in disk evolution models should perhaps be less conservative. Most models that combine disk evolution with the SI assume that planetesimals form when the model predicts $\rhop/\rhog > 1$ at the midplane \citep[e.g.,][]{Carrera_2017,Drazkowska_2017,Coleman_2021}.
\citet{Carrera_2017} had already shown that this rule is much more stringent than the limits of \citet{Carrera_2015}, but that was considered prudent since the SI was less well understood at that time. As a point of reference, for a disk with $\alpha = 10^{-4}$ the $\rhop/\rhog > 1$ criterion would correspond to $Z \approx 0.028$ for the $\tau = 0.12$ particles in our experiment. By comparison, the criterion of \citet{Carrera_2015} gives $Z \approx 0.014$. Since then, \citet{Yang_2017} and \citet{Li_2021} have progressively reduced the critical $Z$ for the SI, and then \citet{Carrera_2021} showed that in a large simulation domain particle filaments form at even lower $Z$ (though without any strong clumping unless a pressure bump is present). Now we present a new result that the SI is resilient to weakly reinforced bumps as well. 

The sources of uncertainty in our model are similar to those of previous simulations of the SI. We can only resolve the gravitational field down to the grid scale, and therefore particle clouds cannot collapse down to realistic planetesimal sizes. As a result, our simulations greatly overestimate the rate of pebble accretion and the aerodynamic drag on a planetesimal. In addition, many physical properties of the disk --- the amount of flaring, the shape of the bump, and the source of reinforcement --- are very poorly constrained. Furthermore, in this work all our simulations have a fixed $a = 1$cm particle size in line with \citet{Carrera_2021}'s estimate of the fragmentation limit for our disk model. Yet, other perfectly plausible disk models predict smaller drift-limited particle sizes \citep[e.g.,][]{Drazkowska_2021} that would make the SI less efficient \citep{Carrera_2015}.

Perhaps most importantly for future work, we feel that the potential link between bump reinforcement, pebble drift, and pebble accretion merits further study. Aside from a more realistic source of bump reinforcement, we hope that future works will model a full particle size distribution. Our concern is not so much that the SI may be less efficient when multiple particle sizes are considered (see \citealt{Krapp_2019}) since recent results suggest that this is not a major concern \citep{Zhu_2020,Schaffer_2021}. Rather, we suspect that the pebble drift effect may be much weaker for smaller sizes. Not only do smaller pebbles drift more slowly, but their stronger coupling means that the gas and solids behave more as a single fluid (a colloid).

In any case, that planetesimals form so readily, even for weakly reinforced, low-amplitude pressure bumps is encouraging for the field of planet formation.  Indeed, given the prevalence of such bumps in disks, even at the earliest stages of disk evolution \citep{Alma_2015,Sheehan_2018,Segura-Cox_2020}, what was once considered to be a challenging step in the planet formation process may now be an inevitable outcome of dust-solid interactions in the vicinity of even very weak bumps.

\acknowledgements
We thank the referee for useful feedback that greatly improved the quality of this work. We also thank Anders Johansen and members of his group with whom we had a chance to discuss the results of this investigation. DC, JBS, and KAK acknowledge support from NASA under {\em Emerging Worlds} through grant 80NSSC21K0037.  The numerical simulations and analyses were performed on {\sc Stampede 2} through XSEDE grant TG-AST120062.

%
%
\bibliographystyle{apj}
\bibliography{references}

\end{document}